\documentclass[5p,twocolumn,times,number]{elsarticle}
\usepackage{amssymb}
\usepackage{color}
\usepackage{here}
\begin{document}
\begin{frontmatter}
\title{Performance of newly constructed plastic scintillator barrel 
in the WASA-FRS experiments and evaluation of radiation damage effects on 
multi-pixel photon counter}
\author[1]{Y.~K.~Tanaka\corref{cor1}} 
\ead{yoshiki.tanaka@a.riken.jp}
\author[2,3,4]{R.~Sekiya} 
\author[2,3]{K.~Itahashi\fnref{fn1} } 
\author[5]{H.~Alibrahim Alfaki}
\author[5]{F.~Amjad }
\author[5,6]{M.~Armstrong }
\author[5]{K.-H.~Behr }
\author[7]{J.~Benlliure }
\author[8,9]{Z.~Brencic }
\author[5,10]{T.~Dickel }
\author[5,11]{V.~Drozd }
\author[5]{S.~Dubey\fnref{fn2} }
\author[1]{H.~Ekawa }
\author[12]{S.~Escrig }
\author[7]{M.~Feijoo-Font\'{a}n }
\author[13]{H.~Fujioka }
\author[1,14,15]{Y.~Gao }
\author[5,10]{H.~Geissel\fnref{fn3} }
\author[16]{F.~Goldenbaum }
\author[7]{A.~Gra\~{n}a Gonz\'{a}lez }
\author[5]{E.~Haettner }
\author[11]{M. N.~Harakeh }
\author[1,17]{Y.~He }
\author[5]{H.~Heggen }
\author[5]{C.~Hornung }
\author[5,18]{N.~Hubbard }
\author[2,3]{M.~Iwasaki }
\author[11]{N.~Kalantar-Nayestanaki }
\author[1,19]{A.~Kasagi }
\author[11]{M.~Kavatsyuk }
\author[5]{E.~Kazantseva }
\author[20,21]{A.~Khreptak }
\author[5]{B.~Kindler }
\author[5]{H.~Kollmus }
\author[5]{D.~Kostyleva }
\author[22]{S.~Kraft-Bermuth }
\author[5]{N.~Kurz }
\author[1,14,15]{E.~Liu }
\author[5]{B.~Lommel }
\author[5]{S.~Minami }
\author[23]{D.~J.~Morrissey }
\author[21,24]{P.~Moskal }
\author[5]{I.~Mukha }
\author[1]{M.~Nakagawa }
\author[5]{C.~Nociforo }
\author[15,25,26]{H.~J.~Ong }
\author[5]{S.~Pietri }
\author[5]{S.~Purushothaman }
\author[12]{C.~Rappold }
\author[5]{E.~Rocco }
\author[7,32]{J.~L.~Rodr\'{i}guez-S\'{a}nchez }
\author[5]{P.~Roy \fnref{fn4}}
\author[27]{R.~Ruber }
\author[1,5,17]{T. R.~Saito }
\author[5]{S.~Schadmand }
\author[5,10]{C.~Scheidenberger }
\author[5]{P.~Schwarz }
\author[16]{V.~Serdyuk }
\author[21,24]{M.~Skurzok }
\author[5]{B.~Streicher }
\author[5,28]{K.~Suzuki }
\author[5]{B.~Szczepanczyk }
\author[14]{X.~Tang }
\author[5]{N.~Tortorelli }
\author[8]{M.~Vencelj }
\author[5]{T.~Weber }
\author[5]{H.~Weick }
\author[5]{M.~Will }
\author[5]{K.~Wimmer }
\author[29]{A.~Yamamoto }
\author[1,30]{A.~Yanai }
\author[5,31]{J.~Zhao } 
\author{for WASA-FRS/Super-FRS Experiment Collaboration}
\cortext[cor1]{Corresponding author}

\address[1]{High Energy Nuclear Physics Laboratory, RIKEN Cluster for Pioneering Research, RIKEN, 351-0198, Wako, Saitama, Japan}
\address[2]{Meson Science Laboratory, RIKEN Cluster for Pioneering Research, RIKEN, 351-0198, Wako, Saitama, Japan}
\address[3]{Nishina Center for Accelerator-Based Science, RIKEN, 351-0198, Wako, Saitama, Japan}
\address[4]{Kyoto University, 606-8502, Kyoto, Japan}
\address[5]{GSI Helmholtzzentrum f\"{u}r Schwerionenforschung GmbH, 64291, Darmstadt, Germany}
\address[6]{Institut f\"{u}r Kernphysik, Universit\"{a}t K\"{o}ln, 50923, K\"{o}ln, Germany}
\address[7]{IGFAE, Universidade de Santiago de Compostela, 15782, Santiago de Compostela, Spain}
\address[8]{Jo\v{z}ef Stefan Institute, 1000, Ljubljana, Slovenia}
\address[9]{University of Ljubljana, 1000, Ljubljana, Slovenia}
\address[10]{Universit\"{a}t Gie\ss{}en, 35392, Gie{\ss}en, Germany}
\address[11]{ESRIG, University of Groningen, 9747 AA, Groningen, The Netherlands}
\address[12]{Instituto de Estructura de la Materia - CSIC, 28006, Madrid, Spain}
\address[13]{Institute of Science Tokyo, 152-8551, Tokyo, Japan}
\address[14]{Institute of Modern Physics, Chinese Academy of Sciences, 730000, Lanzhou, China}
\address[15]{School of Nuclear Science and Technology, University of Chinese Academy of Sciences, 100049, Beijing, China}
\address[16]{Institut f\"{u}r Kernphysik, Forschungszentrum J\"{u}lich, 52425, J\"{u}lich, Germany}
\address[17]{Lanzhou University, 730000, Lanzhou, China}
\address[18]{Institut f\"{u}r Kernphysik, Technische Universit\"{a}t Darmstadt, 64289, Darmstadt, Germany}
\address[19]{Graduate School of Engineering,  Gifu University, 501-1193, Gifu, Japan}
\address[20]{INFN, Laboratori Nazionali di Frascati, 00044, Frascati, Roma, Italy} 
\address[21]{Institute of Physics, Jagiellonian University, 30-348, Krak\'{o}w, Poland}
\address[22]{TH Mittelhessen University of Applied Sciences, 35390, Gie\ss{}en, Germany}
\address[23]{National Superconducting Cyclotron Laboratory, Michigan State University, MI 48824, East Lansing, USA}
\address[24]{Center for Theranostics, Jagiellonian University, 30-348, Krak\'{o}w, Poland}
\address[25]{Joint Department for Nuclear Physics, Lanzhou University and Institute of Modern Physics, Chinese Academy of Sciences, 730000, Lanzhou, China}
\address[26]{Research Center for Nuclear Physics, Osaka University, 567-0047, Osaka, Japan}
\address[32]{CITENI, Campus Industrial de Ferrol, Universidade da Coru\~{n}a, 15403, Ferrol, Spain}
\address[27]{Uppsala University, 75220, Uppsala, Sweden}
\address[28]{Ruhr-Universit\"{a}t Bochum, Institut f\"{u}r  Experimentalphysik I, 44780, Bochum, Germany}
\address[29]{KEK, 305-0801, Tsukuba, Ibaraki, Japan}
\address[30]{Saitama University, 338-8570, Saitama, Japan}
\address[31]{Peking University, 100871, Beijing, China}
                 
\fntext[fn1]{Present address: Department of Physics, The University of Osaka, 560-0043, Osaka, Japan}
\fntext[fn2]{Present address: Auburn University, Auburn, AL 36832, USA}
\fntext[fn3]{Deceased}
\fntext[fn4]{Present address: Variable Energy Cyclotron Centre, Kolkata-700064, India}

\begin{abstract}
A barrel-shaped plastic scintillation counter
with Multi-Pixel Photon Counter (MPPC) readout
has been developed and operated in the first
WASA-FRS experimental campaign at GSI.
The detector was used to measure charged particles emitted from
reactions induced by a 2.5~GeV proton beam incident on a carbon target,
providing particle identification in combination with momentum reconstruction
in a 1~T magnetic field.
The performance of this detector, particularly its response to energy deposition and time resolution,
was systematically investigated as a function 
of count rate and total number of irradiating protons.
A time resolution of $45$--$75$~ps ($\sigma$), depending on the energy deposition, was achieved.
Stable performance was maintained under high-rate conditions up to 1.35~MHz per single counter,
with no significant degradation in either signal amplitude or timing response.
Radiation-induced damage to the MPPCs was observed primarily as a reduction in signal amplitude,
with approximately 35\% decrease 
at an estimated 1~MeV neutron-equivalent 
fluence of $2.4 \times 10^{10}$~cm$^{-2}$. 
\end{abstract}
\end{frontmatter}

\section{Introduction} \label{section_introduction}

In recent years, silicon photomultipliers, also referred to as Multi-Pixel Photon Counters (MPPCs),
have been widely adopted as readout detectors for scintillation counters
in particle- and nuclear-physics experiments.
Their advantages lie in their compactness, lower cost, and insensitivity to magnetic fields,
compared to conventional readout systems using photomultiplier tubes.
These features make MPPCs suitable readout devices for plastic scintillators
as part of particle identification detectors within spectrometer systems,
where the plastic materials are placed in strong magnetic fields and often in limited spaces.
High time resolution can be achieved by coupling MPPCs with fast-timing plastic scintillators, owing to
the intrinsically fast timing response of MPPCs \cite{Cattaneo_IEEE_2014, Cattaneo_NIMA_2016, Stoykov_NIMA_2012, 
Nishimura_NIMA_2020, Korzenev_JPS_2019, Alici_JINST_2018, Onda_NIMA_2019}.

However, one of the potential issues with MPPC readout is its limited radiation tolerance.
It has been reported that radiation damage in silicon sensors 
leads to an increase of leakage currents, higher dark count rates, and a reduction in signal amplitude 
under exposure to various types of irradiation~\cite{Garutti2019_NIMA, Qiang2013_NIMA, Mikhaylov2020, Heering2016_NIMA, Ieki2023_NIMA, Sanchez2009_NIMA, Garutti_arxiv_2017, Barnyakov2016_NIMA}.
These effects can become critical, particularly in experiments involving hadronic beams, where high neutron fluences are expected within relatively short time periods.
Therefore, it is of particular importance to investigate and characterize the performance not only of the sensors themselves, but also of the full MPPC-based scintillator system, in terms of particle reconstruction capabilities under realistic experimental conditions, when it is integrated within a full spectrometer setup.

We developed a plastic scintillator system based on MPPC readout \cite{Sekiya2022} as 
a part of the Wide-Angle Shower Apparatus (WASA) spectrometer system \cite{WASA1, WASA2} and
operated it in the first series 
of the WASA-FRS experiments at GSI \cite{Tanaka2023, Saito2023_EMIS2022}.
The detector was employed for the identification of charged particles emitted from proton- and heavy-ion-induced 
reactions, by measuring their time of flight and the energy deposition 
in combination with momentum reconstruction performed
using tracking detectors inside a 1~T solenoidal magnetic field. 
Two experiments were carried out in the first experimental campaign, one on the spectroscopy of  $\eta^\prime$-mesic nuclei \cite{Tanaka2019INPC} and the other on light hypernuclear spectroscopy using heavy-ion-induced reactions \cite{Saito2021-NatRev}.
 
In this article, we report on a systematic investigation of 
the performance of the newly constructed plastic scintillator barrel 
from the analysis of the spectroscopy experiment of  $\eta^\prime$-mesic nuclei.
The detector response to energy deposition and the time resolution were analyzed under various conditions, including dependence on the counting rate and the total number of irradiating protons for evaluating radiation tolerance.
The structure of this paper is as follows.
First, the experimental setup and measurements are introduced in Section~\ref{sec_experiment}.
The data-analysis procedure is presented in Section~\ref{sec_analysis}, followed by results and discussions 
in Section~\ref{sec_results}.
Finally, the conclusions are summarized in Section~\ref{sec_conclusion}.

 
\section{Experiment} \label{sec_experiment}

\begin{figure}[hbt]
\centering
\includegraphics[scale=0.80]{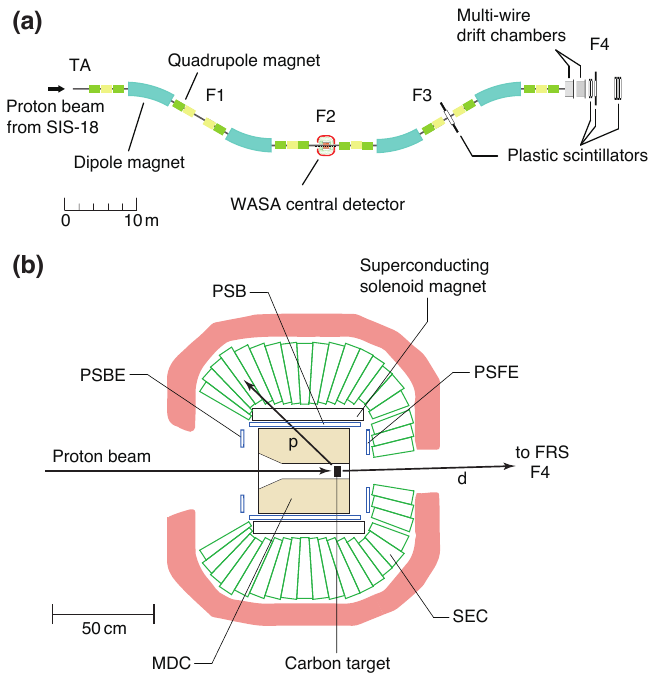}
\caption{
(a) A schematic experimental setup with the WASA central detector and the FRS at GSI. 
A 2.5 GeV proton beam impinged on a carbon target at F2. 
Deuterons emitted in the ($p$,$d$) reaction were analyzed by the F2–F4 section of FRS. 
Decay particles from $\eta^\prime$-mesic nuclei were detected by the WASA detector at F2.
(b) A schematic configuration of the WASA central detector at F2.
A carbon target was installed at 15~cm downstream position from the central point of the WASA detector.
Emitted particles from reactions were measured by the MDC, PSB, PSFE, PSBE, and SEC detectors. 
See text for details of these detectors.}
\label{fig:setup_wasa_frs}
\end{figure}

The experimental setup for the spectroscopy of  $\eta^\prime$-mesic nuclei is illustrated in Figure~\ref{fig:setup_wasa_frs}. 
The central part of the WASA detector   
was installed at the F2 focal plane of 
the fragment separator FRS \cite{FRS_Geissel_92}. 
We employed a 2.5~GeV proton beam extracted from the SIS-18 synchrotron,
with a spill length of 10~s. The beam impinged on a carbon target 
placed at F2 to produce $\eta^\prime$-mesic nuclei 
with the ${}^{12}$C($p$,$d$)$\eta^\prime \otimes {}^{11}$C reaction.
The typical beam rate was $\sim 5 \times 10^{8}$/s. 
The carbon target had an areal density of 4~g/cm$^2$ along the beam axis
and was installed inside the WASA detector at a position 15~cm downstream of the detector's central point.

The F2--F4 section of the FRS was operated 
as a high-resolution momentum spectrometer 
at a magnetic rigidity of 9.4~Tm.
Forward-emitted deuterons from the ${}^{12}$C($p$,$d$) reaction 
near the $\eta^\prime$-meson production threshold
were identified by time-of-flight measurements 
between the plastic scintillators installed at F3 and F4. 
Their momenta were reconstructed from trajectories 
measured with the multi-wire drift chambers at F4,
yielding a momentum resolution of $\sigma_P / P \sim 1/3000 $,
and were then used to calculate the missing mass
of the ($p$,$d$) reaction.
Further details on the particle identification at the FRS and on the tracking detectors can be found in Refs.~\cite{Tanaka2023, Tanaka2018}.

The WASA central detector was used to measure and tag particles
emitted from the decay of the $\eta^\prime$-mesic nuclei, 
in order to enhance the signal-to-background ratio of the spectrum 
\cite{Ikano_arxiv_2024} compared to the previous experiment~\cite{Tanaka2018, Tanaka2016}.  
As shown in the lower panel of Figure~\ref{fig:setup_wasa_frs}, 
the WASA detector is a spectrometer system consisting of several components: 
a mini-drift chamber (MDC) \cite{MDC_PhD_MAREK_JACEWICZ},
a plastic scintillator barrel (PSB) 
and forward and backward end caps (PSFE and PSBE, respectively), 
a superconducting solenoid magnet \cite{Ruber_2003}, 
and a scintillator electromagnetic 
calorimeter (SEC) \cite{CsI_PhD_INKEN_KOCH}.  
Momenta of charged particles were reconstructed 
from their trajectories in a magnetic field of 1~T, 
measured by the MDC.  
The plastic scintillators provided timing and energy-deposition measurements, and contributed to particle identification 
in combination with the momentum information from the MDC.  
The SEC was used to detect high-energy photons emitted from the decay of neutral mesons.

\begin{figure}[hbt]
\centering
\includegraphics[scale=0.075]{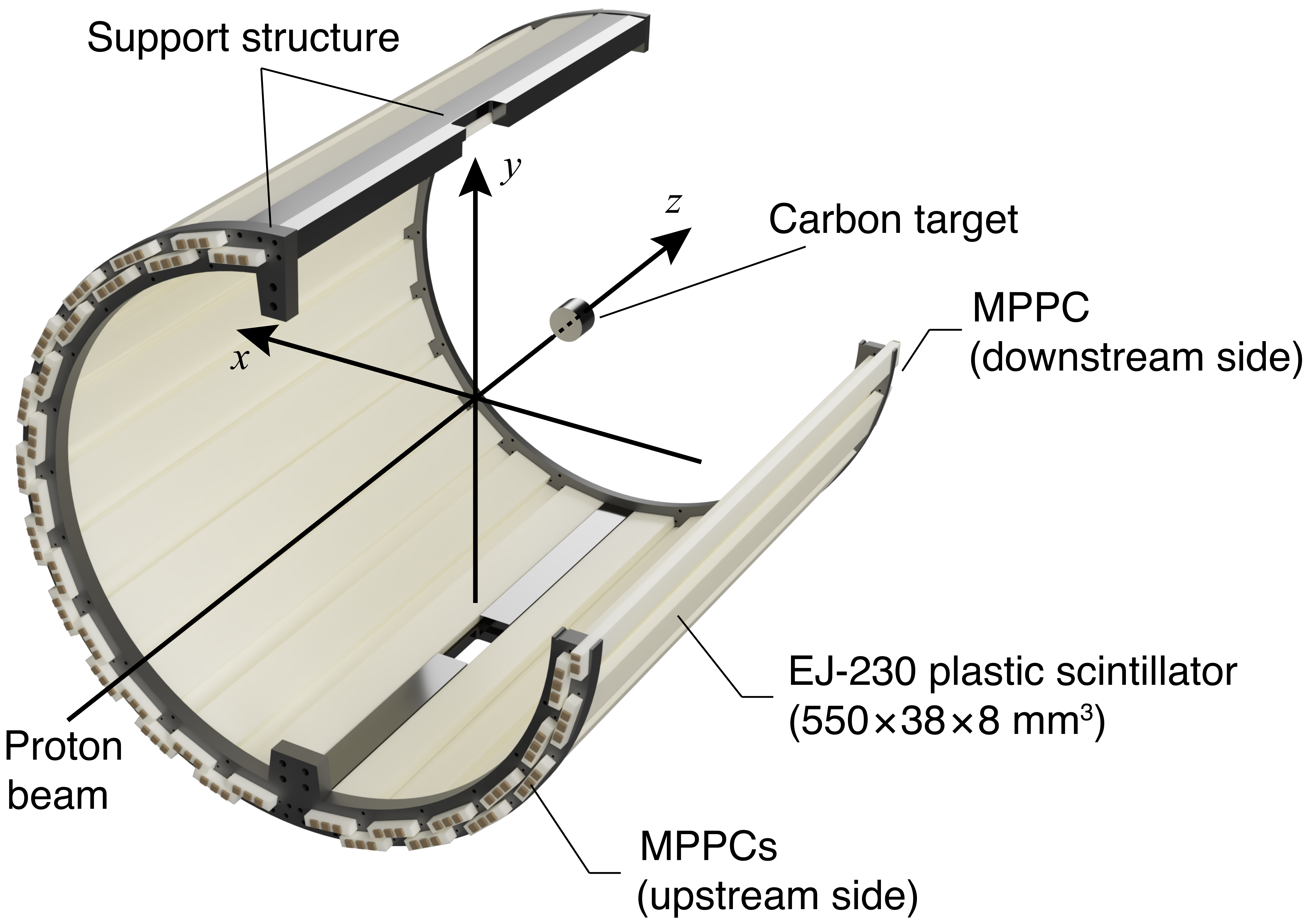}
\caption{A three-quarter section view of the PSB and the location of the carbon target. 
46 slats of plastic scintillator (EJ-230) with dimensions of $550 \times 38 \times 8$~mm$^3$ were alternately positioned at radial distances 
of 221~mm and 232~mm from the central beam axis, 
with overlapping regions between adjacent bars. 
Three MPPCs (S13360-6050PE) were attached to both the upstream and downstream ends 
of the scintillator bar with respect to the beam direction.
The center of the plastic bar was positioned at $z=0$~mm,
while the target was installed at $z=150$~mm.}
\label{fig:psb_full_view}
\end{figure}

In this article, we focus on the analysis 
of the PSB detector, which was newly developed and constructed 
for the present experiment~\cite{Sekiya2022}.  
A schematic view of the PSB is shown in Figure~\ref{fig:psb_full_view}.  
The detector consists of 46 plastic scintillation bars, 
each with dimensions of $550 \times 38 \times 8$~mm$^3$, 
arranged in a cylindrical barrel configuration.  
The bars were alternately positioned at radial distances 
of 221~mm and 232~mm from the central beam axis, 
with overlapping regions between adjacent bars.  
The entire azimuthal angle range was covered, except for regions around 
$\phi = \pi/2$ and $3\pi/2$ due to the support structure.  
Eljen Technology EJ-230 was used as the scintillator material, 
which has an attenuation length of 120~cm and rise 
and decay times of 0.5~ns and 1.5~ns, respectively. 

The MPPC S13360-6050PE manufactured by Hamamatsu Photonics 
was employed as the photon detector. 
It has a photosensitive area of $6 \times 6$~mm$^2$ and a pixel pitch of 50~$\mu$m.  
Three MPPCs were electrically connected in series on a printed circuit board and directly 
attached to both the upstream and downstream ends of the scintillator bar with respect to the beam direction, covering approximately 36\% of the readout surface area.  
An optical grease TSK5353 (Momentive Performance Materials) was used 
to optically couple the plastic bars to the MPPCs.
Each MPPC was operated at a bias voltage of 55.0~V, corresponding to an overvoltage of 3.5~V. 
The bias voltage was supplied using 
a regulated power supply (PMX250-0.25A)
manufactured by Kikusui Electronics.

Raw signals from the MPPCs were transmitted 
via 7~m-long CLF100 coaxial cables 
to amplifier modules based on the design reported in Ref.~\cite{Cattaneo_IEEE_2014}.  
Two modifications were introduced to meet the requirements 
of the present experiment.  
First, the total resistance of the low-pass filter circuit 
used for feeding the bias voltage 
was reduced from 3.6~k$\Omega$ to 
110~$\Omega$ to 
improve high-rate performance.
Second, the $\Pi$-type attenuator was modified 
to match the expected signal amplitudes to the dynamic range 
of the subsequent readout electronics.
The amplified signals were digitized 
using a CAEN V1742 waveform digitizer operating at a sampling rate of 2.5~GHz 
for timing and amplitude analysis.
Split  signals were also sent to a constant-fraction discriminator (MCFD-16, Mesytec GmbH), 
and the rate information was recorded using a 250~MHz scaler (CAEN V830).

The intensity of the primary proton beam was continuously monitored 
during the experiment using a 
Secondary Electron Transmission 
Monitor (SEETRAM)~\cite{Jurado_NIMA_2002_SEETRAM}, 
installed at the standard target area (TA) of the FRS.  
The current signal from the SEETRAM was converted into a pulse frequency 
using a current digitizer (GSI CD1011) and recorded with the CAEN V830 scaler.  
The SEETRAM response was calibrated for 2.5~GeV protons 
at the end of this experiment, 
with an accuracy of approximately 5\%.

The data-acquisition system was triggered by several conditions.
The primary trigger was generated 
by a time-of-flight-based coincidence of 
scintillator signals from F3 and F4 of the FRS, 
enabling efficient recording of events associated with the ($p$,$d$) reaction.  
In addition, a downscaled signal from the F4 scintillator alone, 
with a factor of 128, was added to the trigger logic to record a fraction of the ($p$,$p'$) reaction events. 
Downscaled signals from the PSB, PSBE, and PSFE detectors, with factors of $2^{13}$--$2^{14}$, as well as a 5 Hz clock signal, were also included for calibration purposes.
In the following sections, data collected under all trigger conditions were combined and analyzed.

 
\section{Data analysis} \label{sec_analysis}

Waveform data of the PSB signals were analyzed 
to extract the hit timing and energy deposition of charged particles.
A software-based method of 
a constant-fraction discriminator \cite{Codino_NIMA_2000_CFD}
was employed to define the hit timing 
while suppressing time-walk effects.
We adopted a delay parameter of $2.8$~ns and a fraction parameter of 
$0.4$ by optimizing the resulting time resolution.
The energy deposition was evaluated by integrating 
baseline-subtracted waveform 
within a time window of [$-4$, $20$] ns relative 
to the point where the signal crossed a fixed threshold of $-12$~mV.
Since this procedure emulates the function of 
a charge-to-digital converter (QDC), 
the resulting quantity is hereafter referred to as the QDC value.  
These analyses were performed for the MPPC readouts at both the upstream and downstream ends of each scintillator bar.  
The hit timing of the PSB slat was then determined as the arithmetic mean of the timings from both ends, while the energy deposition was calculated as the geometric mean of the two QDC values, in order to eliminate dependence on the longitudinal hit position ($z$).

The momentum and trajectory of charged particles were reconstructed 
by fitting drift-length data measured by the MDC.
First, an elastic-arm algorithm \cite{Ohlsson_CompPhys_1992_EAA} 
was applied to select 
a combination of hits forming each trajectory.  
The selected hits were then fitted 
by employing a Kalman-filter algorithm with the GENFIT toolkit~\cite{Genfit_NIMA}.
The position resolution of each MDC layer was 
estimated to be $\sim$200~$\mu$m ($\sigma$), 
resulting in a typical momentum resolution of approximately 15\% at 0.5~GeV/$c$.
Details of the MDC analysis will be reported elsewhere~\cite{Sekiya_PhD}.

\begin{figure}[hbt]
\centering
\includegraphics[scale=0.46]{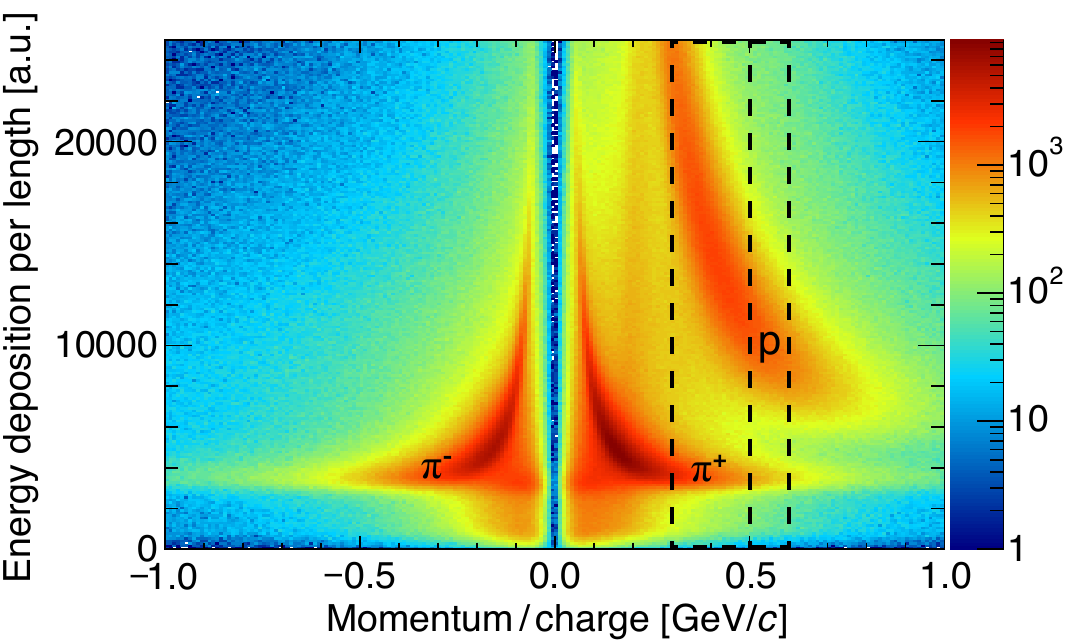}
\caption{A typical example of a particle identification plot. The abscissa shows the reconstructed momentum divided by the sign of the charge, as obtained from the track fitting analysis. The ordinate shows the energy deposition per unit track length in the PSB volume. Boxes with the dashed lines indicate the selected momentum regions used for analyzing the QDC response and time resolution.}
\label{fig:particle_identification}
\end{figure}

Charged particles were identified by combining the reconstructed momentum and the
measured energy deposition.
Figure~\ref{fig:particle_identification} shows an example of a particle identification plot.
The horizontal axis represents the reconstructed momentum divided by the charge,
while the vertical axis shows the energy deposition divided by the track length within the PSB volume, as evaluated from the track fitting analysis.
Data from all PSB slats were combined after applying individual time offset and gain corrections.
Protons ($p$) and charged pions ($\pi^{\pm}$) are clearly distinguished, as demonstrated in the figure.

The response to the energy deposition was analyzed using QDC spectra obtained from the individual upstream and downstream MPPCs, as well as from their geometric mean.
Two momentum regions were selected for this analysis: 0.3--0.5~GeV/$c$ and 
0.5--0.6~GeV/$c$, as illustrated in Figure~\ref{fig:particle_identification}.
The former was used to analyze positive pions corresponding to 
minimum ionizing particles, while the latter was for protons  
with approximately 3 times higher energy depositions.
We further selected the longitudinal hit position within 100~mm $\leq z \leq 200$~mm
for the energy-deposition analysis
in order to reduce the effect of light attenuation along the $z$-direction 
in the QDC spectra of the individual upstream and downstream MPPCs.

\begin{figure}[hbt]
\centering
\includegraphics[scale=0.46]{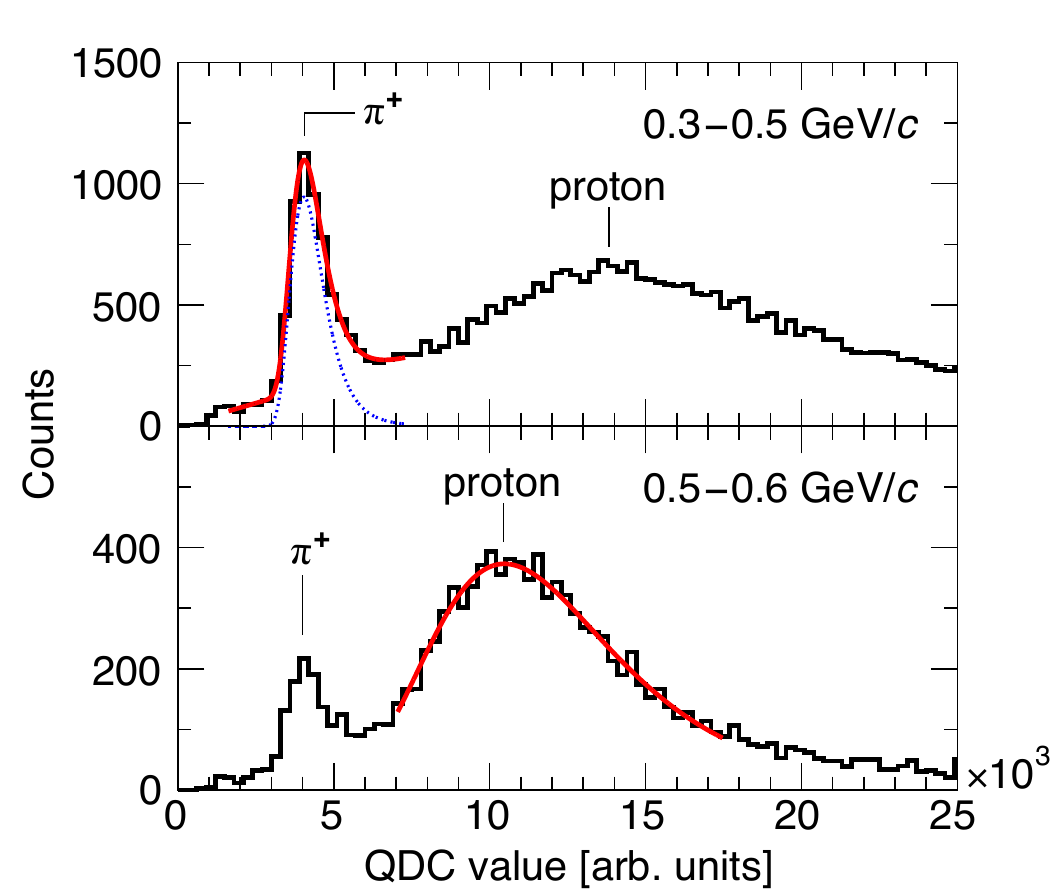}
\caption{ Examples of QDC spectra for the momentum ranges of 0.3--0.5~GeV/$c$ (upper panel) and
 0.5--0.6~GeV/$c$ (lower panel). The geometric mean of the QDC values from the upstream 
 and downstream MPPCs was plotted.
The red solid curves represent fit functions (see text for details). 
The blue dotted curve in the upper panel shows the peak component of the fit function.}
\label{fig:qdc_spectra_example}
\end{figure}

Figure~\ref{fig:qdc_spectra_example} shows examples of 
the QDC spectra for one of the PSB slats. 
The $\pi^{+}$ peak
observed in the momentum range of 0.3--0.5~GeV/$c$ was fitted with an empirical function of the form $f(x) = p_0 \exp{(-(x-p_1)^2/(p_2 +p_3 x)^2 )} + (p_4 + p_5 x)$, where the first term represents an asymmetric peak structure, and the second term accounts for the continuous background arising from the tail of the proton peak. In contrast, the proton peak in the 0.5--0.6~GeV/$c$ range was well reproduced using only the first term, as the pion contribution in this region was sufficiently small. 
The peak position and the full width at half maximum (FWHM) were extracted 
to characterize the PSB response in terms of energy deposition.

The time resolution of the PSB 
was evaluated for the inner slats by using tracks 
penetrating the overlapping region between the inner and outer PSB slats.
Two data sets, one for $\pi^{+}$ in the momentum range of 0.3--0.5~GeV/$c$ and the other for protons in 0.5--0.6~GeV/$c$, were selected 
based on the momentum and QDC values.
In addition, the longitudinal hit position in a range of $-25$~mm $\leq z \leq$ 25~mm
was selected, except when scanning the $z$-dependence of the resolution.
 
We defined the following three quantities for evaluating the time resolution:
\begin{eqnarray}
T_\mathrm{1UD} &=& T_\mathrm{1U} - T_\mathrm{1D} + f \cdot z \label{eqn_tresol_1}\\
T_\mathrm{1U2} &=& T_2 
- T_\mathrm{1U} - \frac{f\cdot z}{2} \label{eqn_tresol_2} \\
T_\mathrm{1D2} &=& T_2  
- T_\mathrm{1D} + \frac{f\cdot z}{2} \label{eqn_tresol_3}.
\end{eqnarray}
Here, $T_2 = (T_\mathrm{2U}+T_\mathrm{2D})/2$, 
and
$T_{ij}$ ($i=1,2$ and $j=\mathrm{U,D}$) represents the hit 
timing measured at the upstream ($j=\mathrm{U}$) or downstream ($j=\mathrm{D}$) MPPCs 
of the inner ($i=1$) or outer ($i=2$) PSB slat.
$z$ denotes the longitudinal hit position at the inner PSB slat, obtained from track fitting. 
A factor $f$ was introduced to compensate for the $z$-position dependence of 
the defined three quantities and was determined to
 be $f=14.8 \pm 0.1$~ps/mm. 
Each of the distributions defined by Equations~(\ref{eqn_tresol_1})--(\ref{eqn_tresol_3}) exhibits a peak structure. The width of each peak ($\sigma$) was extracted by fitting it with a Gaussian function.
All combinations of the two overlapping PSB slats were used for this analysis, 
with individual time offsets corrected.

The obtained widths ($\sigma_\mathrm{1UD}$, $\sigma_\mathrm{1U2}$, and 
$\sigma_\mathrm{1D2}$) can be related to 
the individual time resolutions of $T_\mathrm{1U}$, $T_\mathrm{1D}$, and 
$T_\mathrm{2}$ ($\sigma_\mathrm{1U}$, $\sigma_\mathrm{1D}$, and $\sigma_\mathrm{2}$, respectively) by 
the following equations:
\begin{eqnarray}
\sigma^2_\mathrm{1UD} &=& \sigma^2_\mathrm{1U} +  \sigma^2_\mathrm{1D} + f^2 \sigma^2_{z} \label{eqn_tresol_4}\\
\sigma^2_\mathrm{1U2} &=& \sigma^2_\mathrm{1U} +  \sigma^2_\mathrm{2} + \frac{f^2 \sigma^2_{z}}{4} \label{eqn_tresol_5} \\
\sigma^2_\mathrm{1D2} &=& \sigma^2_\mathrm{1D} +  \sigma^2_\mathrm{2} + \frac{f^2 \sigma^2_{z}}{4} \label{eqn_tresol_6}.
\end{eqnarray}
Note that $\sigma_{z}$ can be obtained from the track fitting analysis. 
Therefore, Equations (\ref{eqn_tresol_4})--(\ref{eqn_tresol_6}) can be solved to determine $\sigma_\mathrm{1U}$, $\sigma_\mathrm{1D}$, and $\sigma_\mathrm{2}$. The time resolution of the inner slat $\sigma_1 $ can be obtained as well by $\sigma_1 = \sqrt{\sigma_\mathrm{1U}^2 + \sigma_\mathrm{1D}^2 } / 2$, since the hit time at the inner slat is defined as $T_{1} = (T_\mathrm{1U}+T_\mathrm{1D})/2$.

The QDC response and time resolution were systematically analyzed
as functions of both the count rate and the total number of irradiating protons.
The count-rate dependence was studied to investigate performance under high-rate conditions, 
using scaler information within a time window of [$-100$, 0]~ms relative to each hit.
In contrast, the dependence on the total number of incident protons was analyzed 
in order to evaluate possible radiation damage effects.
The total number of protons was determined by integrating the SEETRAM current from the start of 
the production measurements.
In this experiment, a total of $1.1 \times 10^{14}$ protons impinged on the target.
Long-term radiation effects were analyzed by dividing the entire data into 11 subsets, each corresponding to $0.1 \times 10^{14}$ incident protons. 


\section{Results and discussions} \label{sec_results}

\subsection{Results of QDC analysis} \label{subsec_results_qdc}

\begin{figure}[hbt]
\centering
\includegraphics[scale=0.46]{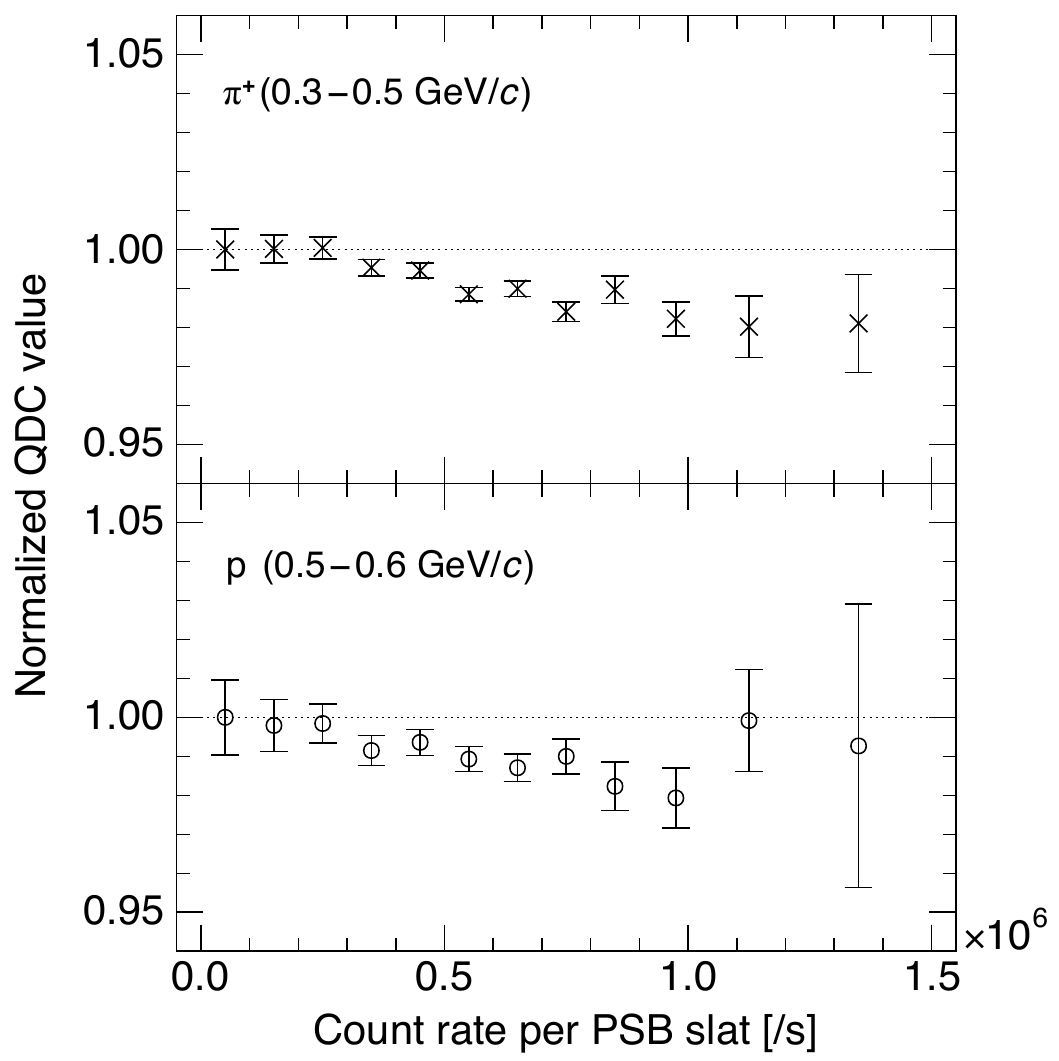}
\caption{Normalized QDC values as a function of the counting rate per individual PSB slat. The geometric mean of the QDC values from the upstream and downstream MPPCs was used. The QDC values were normalized to the first data point, which corresponds to the lowest rate of $5 \times 10^4$~counts/s.}
\label{fig:qdc_rate_dependence}
\end{figure}

Analyzed peak values of the QDC are presented in Figure~\ref{fig:qdc_rate_dependence} to illustrate 
the dependence on the count rate. Data sets corresponding to $\sim 20\%$ of the total incident protons
were used in the analysis of the rate dependence.
The QDC values are normalized to those obtained at the lowest count 
rate for each of the data sets: $\pi^{+}$ in the 0.3--0.5~GeV$/c$ range
and proton in the 0.5--0.6~GeV$/c$ range. 
No significant reduction greater than 2\% is observed up to a counting rate of $1.35 \times 10^6$/s. 
This result demonstrates a substantial improvement compared to the prototype detector previously reported in Ref.~\cite{Sekiya2022}, 
where a visible drop of approximately $20\%$ was found already at 1~MHz.
The improvement can be attributed to the reduced resistance 
in the low-pass filter circuit as well as the use of a more stable power supply for biasing the MPPCs in the present experiment.

\begin{figure}[H]
\centering
\includegraphics[scale=0.46]{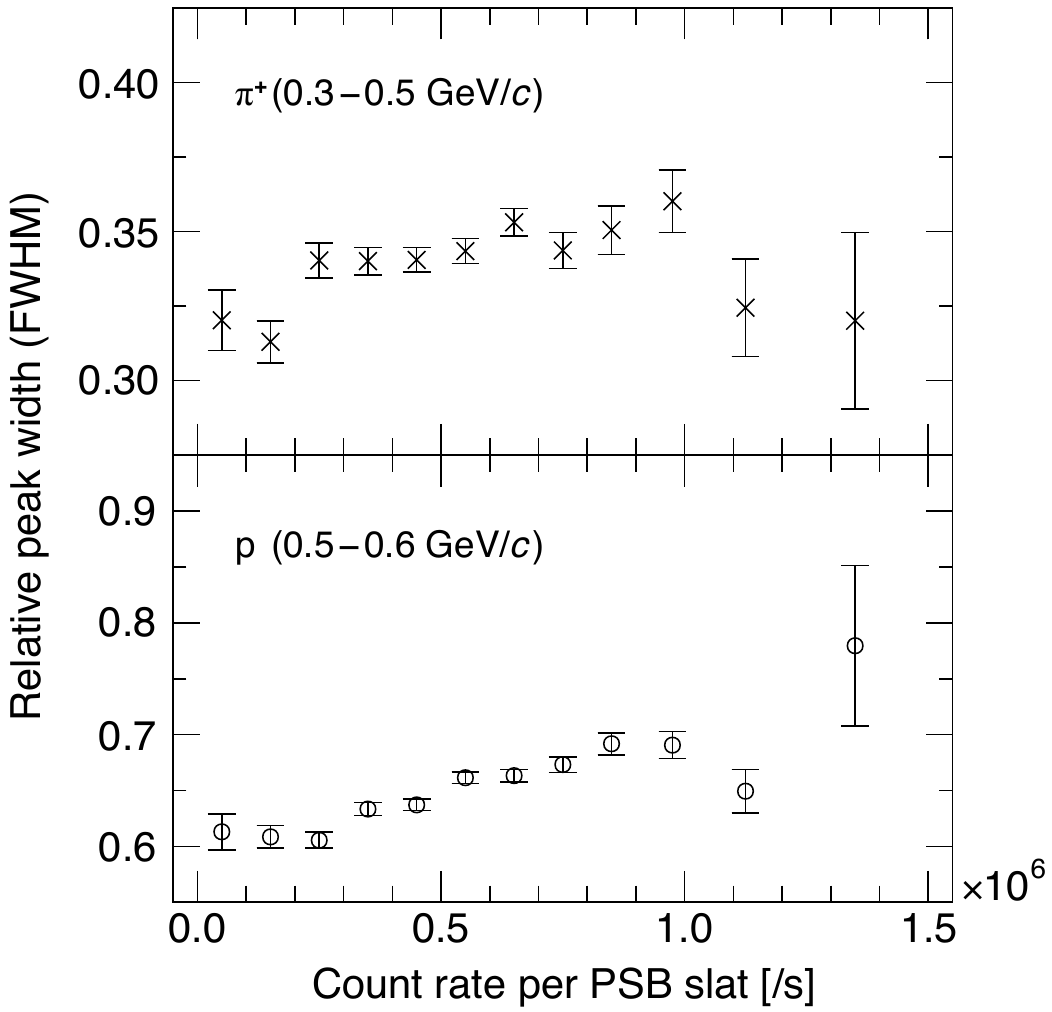}
\caption{Relative peak widths (FWHM) of QDC as a function of the counting rate per individual PSB slat. The geometric mean of the QDC values from the upstream and downstream MPPCs was used.}
\label{fig:width_rate_dependence}
\end{figure}

Obtained widths of the QDC peaks are shown
as a function of the count rate in Figure~\ref{fig:width_rate_dependence}.
The widths are normalized to the corresponding QDC peak values 
under each condition.
The observed widths remain stable over the entire range of count rates,  
taking values in the range of 0.31--0.36 for the $\pi^{+}$ data set and 
 0.61--0.78 for the proton data set.
This also represents a significant improvement compared to the results of the prototype detector \cite{Sekiya2022}.

\begin{figure}[hbt]
\centering
\includegraphics[scale=0.51]{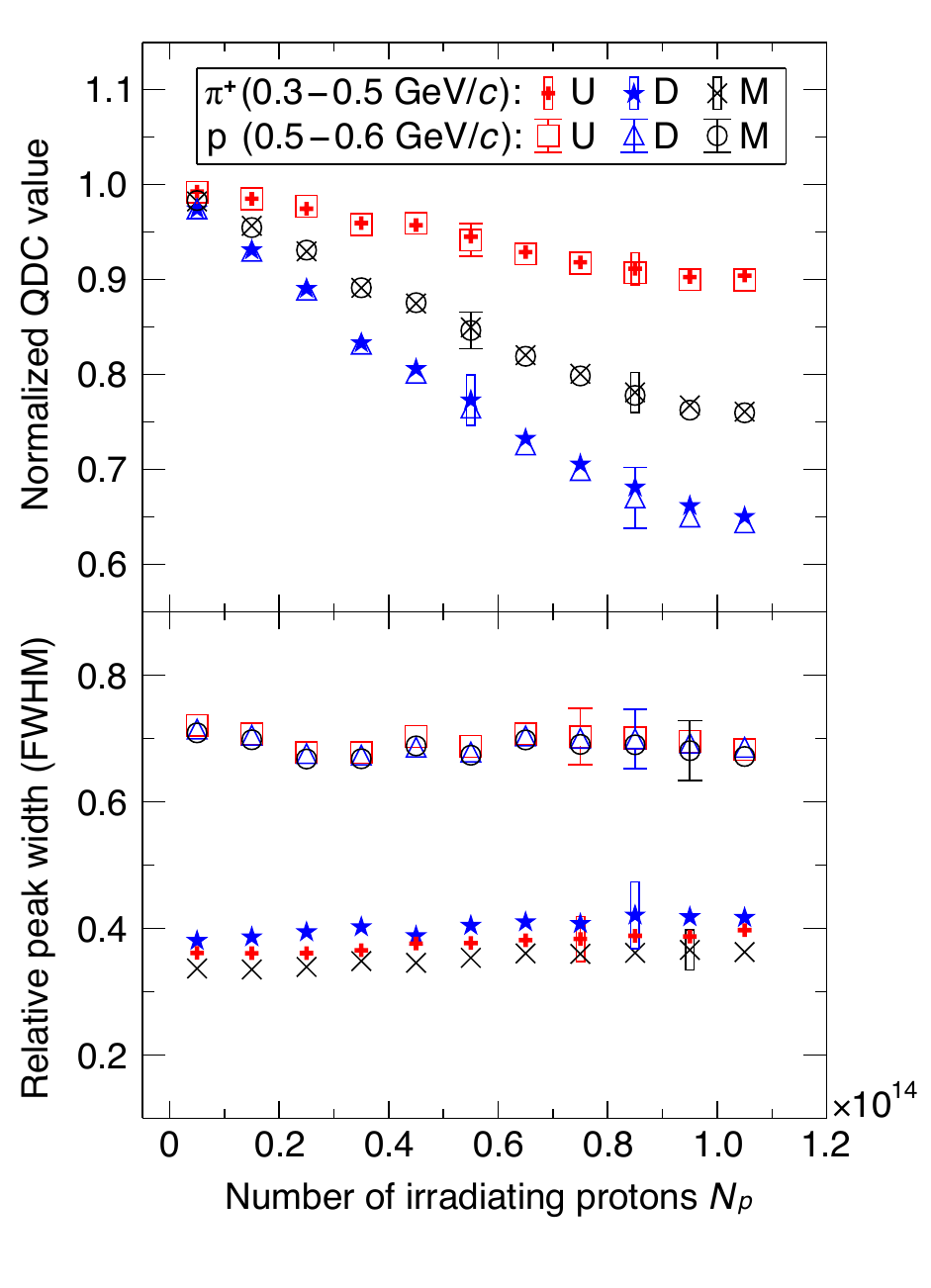}
\caption{Normalized software QDC values (upper panel) and relative peak widths (lower panel) as functions of the total number of irradiating protons, $N_p$. The QDC values were normalized to the extrapolated values at $N_p=0$. Error bars are shown for representative points.}
\label{fig:qdc_reducition_irr_time}
\end{figure}

The long-term trends of the QDC response are shown in Figure~\ref{fig:qdc_reducition_irr_time} 
as functions of the total number of protons $N_p$ impinging on the carbon target.
Each plot of the QDC values in the upper panel 
is normalized to unity at the extrapolated value for $N_p = 0$.
The QDC value based on the geometric mean decreased to 76\%,
primarily due to the QDC of the downstream MPPC which dropped to 63--65\%.
The observed slope of the reduction is the same for the $\pi^{+}$ (0.3--0.5~MeV/$c$) and proton (0.5--0.6~MeV/$c$) data sets.
Since the downstream MPPC is located closer to the reaction target,
this reduction can be attributed to effects due to radiation damage.
In contrast, widths of the QDC peaks remain stable over the entire irradiation time, 
as shown in the lower panel of Figure~\ref{fig:qdc_reducition_irr_time}. 

\begin{figure}[hbt]
\centering
\includegraphics[width=0.84 \linewidth]{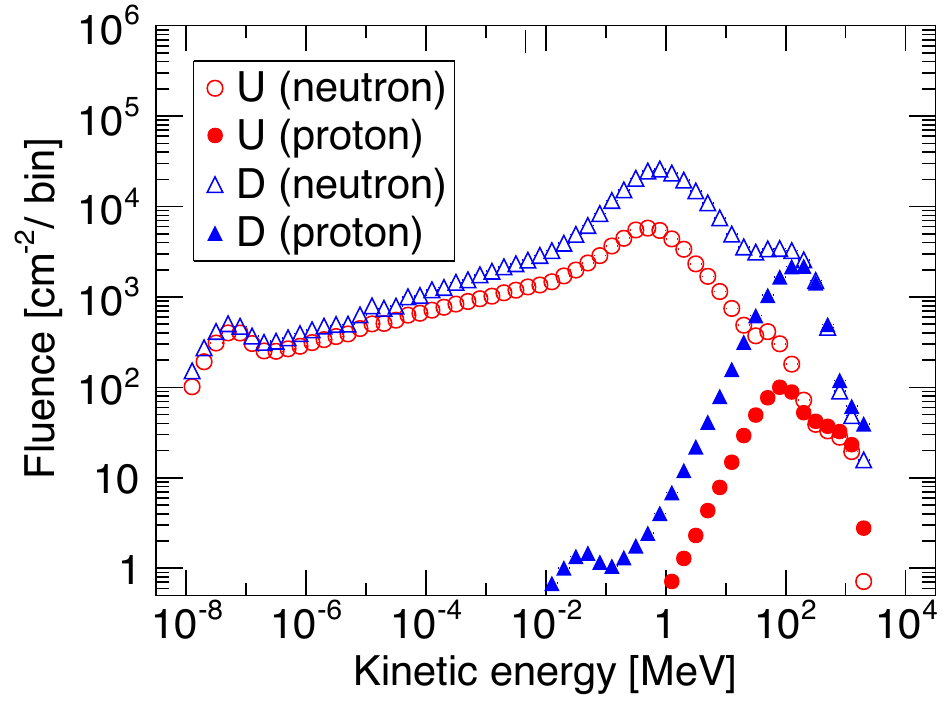}
\caption{Simulated non-weighted fluences of protons and neutrons at the locations of the upstream (U) and downstream (D) MPPCs, assuming $10^9$ incident protons on a 4~g/cm$^2$ carbon target.}
\label{fig:sim_fluence}
\end{figure}

The number of irradiating protons can be translated 
into the 1~MeV neutron-equivalent fluence at each position 
of the upstream and downstream MPPCs. 
For this estimation, a Monte-Carlo simulation was performed 
using the Geant4 framework~\cite{Agostinelli_NIMA_2003_Geant4} to first evaluate
the energy- and particle-dependent fluences at the MPPC locations\footnote{Geant4 version 
10.6.1 was used in the persent simulation.}.
The FTFP\_BERT\_HP physics list of the Geant4 was employed \cite{Geant_NIMA2016}, which
incorporates high-precision neutron models and cross sections at 
low energies. 
The resulting fluences of neutrons and protons 
for $10^9$ incident protons are shown in Figure~\ref{fig:sim_fluence}. 
Contributions of other particles such as $\pi^{\pm}$ 
and $e^{\pm}$ were negligibly small compared to those by the neutrons and protons. 
The obtained spectra of all these particles were then integrated 
with the corresponding NIEL (Non-Ionizing Energy Loss) 
scaling factors in silicon~\cite{Lindstrom_NIMA_2003_NIEL} 
normalized to the value for 1~MeV neutron~\cite{ASTM}.
As a result, the 1~MeV neutron-equivalent fluences per $10^9$ 
incident protons were estimated to be 
$3.8 \times 10^{4}$~cm$^{-2}$ and $2.3 \times 10^{5}$~cm$^{-2}$ 
at the upstream and downstream MPPC positions, respectively.

\begin{figure}[hbt]
\centering
\includegraphics[width=0.88 \linewidth]{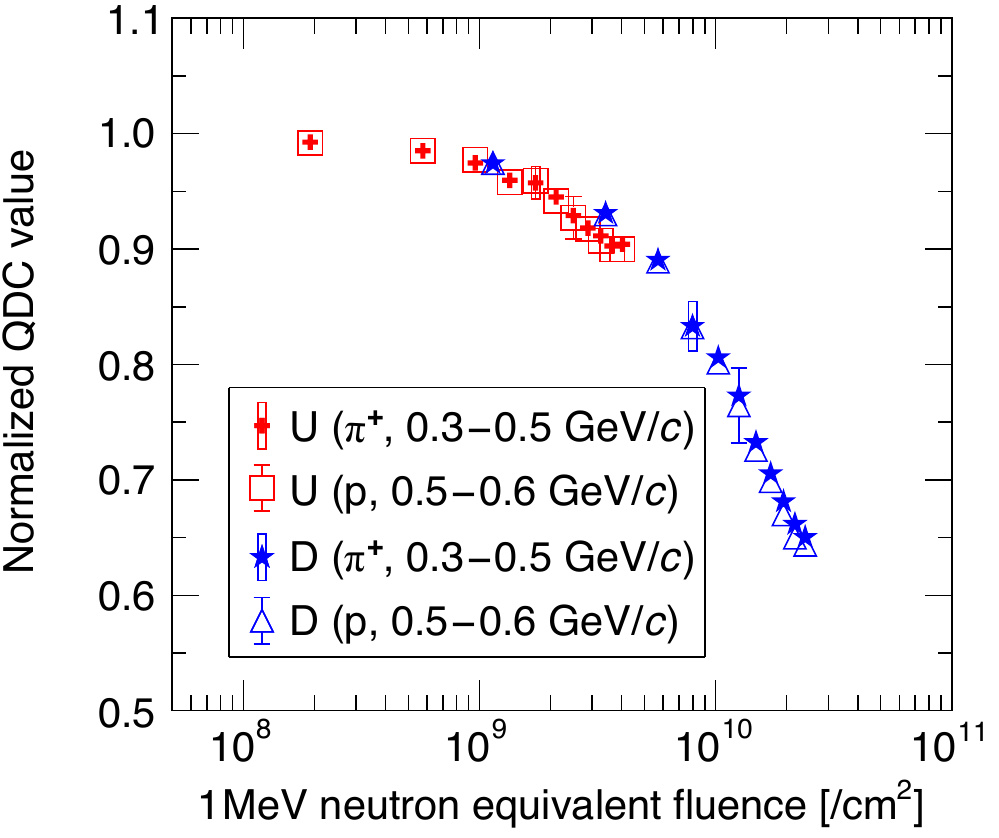}
\caption{Normalized QDC values as a function of the integrated 1 MeV neutron-equivalent fluence in silicon. The QDC values were normalized in the same manner as in Figure~\ref{fig:qdc_reducition_irr_time}.
Error bars are shown for representative points.}
\label{fig:qdc_reduction_fluence}
\end{figure}

Figure~\ref{fig:qdc_reduction_fluence} shows the reduction of the QDC values 
as a function of the 1~MeV neutron-equivalent fluence. 
The reductions observed in the upstream and downstream MPPCs 
follow a consistent trend in terms of the 1~MeV neutron-equivalent fluence,
indicating that the equivalent fluence serves as 
a suitable index to characterize the amplitude reduction of the MPPCs.
The largest reduction observed in the present data is 
35--37\% 
at a total fluence of $2.4 \times 10^{10}$~cm$^{-2}$. 
This level of degradation is consistent in magnitude with results reported 
in Refs.~\cite{Qiang2013_NIMA, Mikhaylov2020}.

\subsection{Results of time-resolution analysis} \label{subsec_time_resolution}

\begin{figure}[hbt]
\centering
\includegraphics[width=0.85 \linewidth]{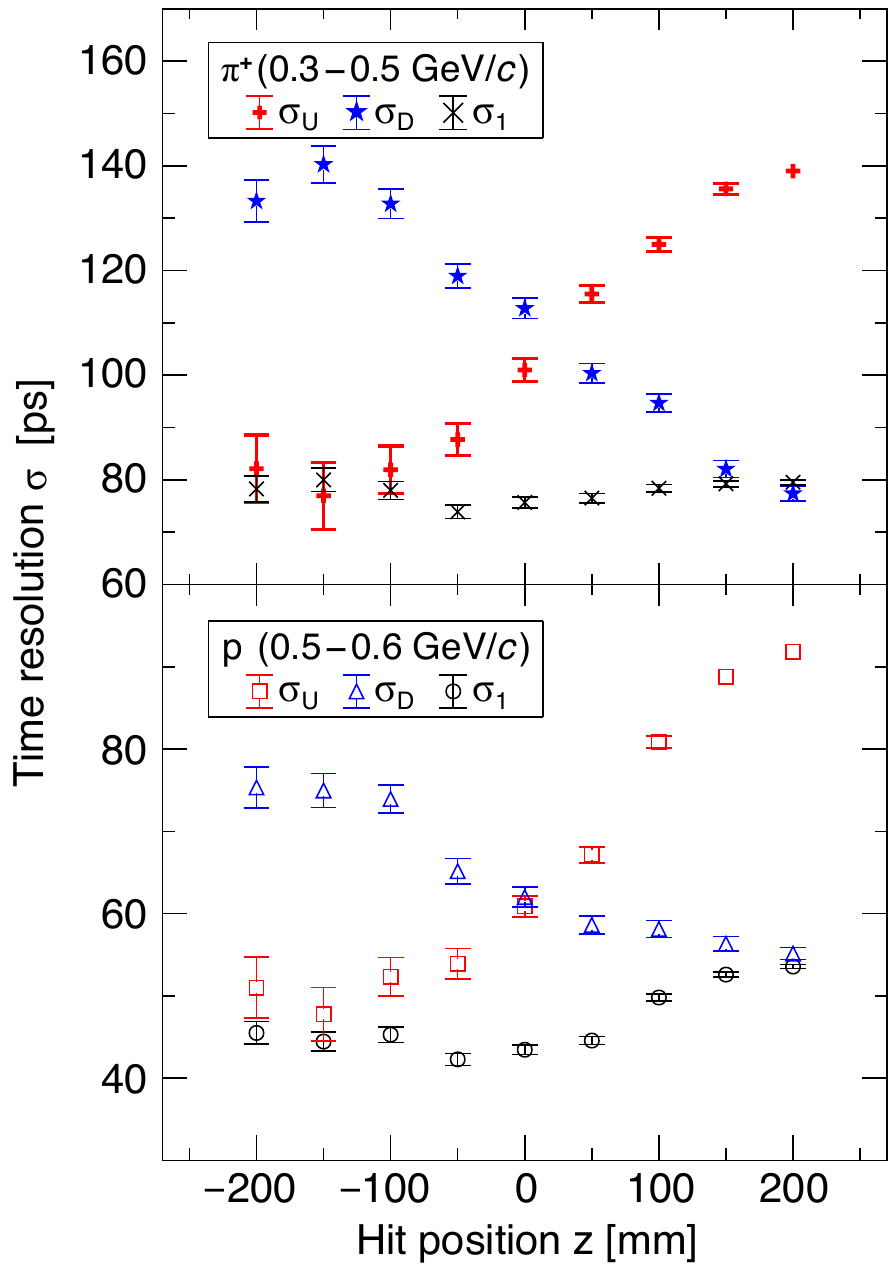}
\caption{ Dependence of the time resolutions ($\sigma_1$, $\sigma_\mathrm{U}$, and $\sigma_\mathrm{D}$)
on the longitudinal hit position $z$ for $\pi^{+}$ in the momentum range of 0.3--0.5 GeV/$c$ (upper panel) and for protons of 0.5--0.6 MeV/$c$ (lower panel).}
\label{fig:time_resolution_position_dependence}
\end{figure}

The analyzed position dependence of the time resolution
is shown in Figure~\ref{fig:time_resolution_position_dependence} 
for $\pi^{+}$ in the momentum range of 0.3--0.5 GeV/$c$ and for protons 
of 0.5--0.6 GeV/$c$.
The achieved time resolutions $\sigma_1$ exhibit a moderate dependence 
on $z$, with values in the ranges of 74--80~ps and 42--54~ps ($\sigma$) 
for the $\pi^{+}$ and proton data, respectively. 
In contrast, the individual contributions from the upstream and downstream MPPCs ($\sigma_\mathrm{U}$ and $\sigma_\mathrm{D}$) 
show a clear dependence on the hit position $z$, which can be understood in terms of light attenuation and differences in photon path length.
 It should be noted that the particle track length in the plastic bar is correlated
 with the hit position, as the target is located at $z=150$~mm
 and the center of the PSB at $z=0$~mm. 
 The observed asymmetry between $z \leq 0$~mm and $0$~mm $\leq z$,
 especially in the proton case, can be understood in this context.
 
\begin{figure}[hbt]
\centering
\includegraphics[width=0.84 \linewidth]{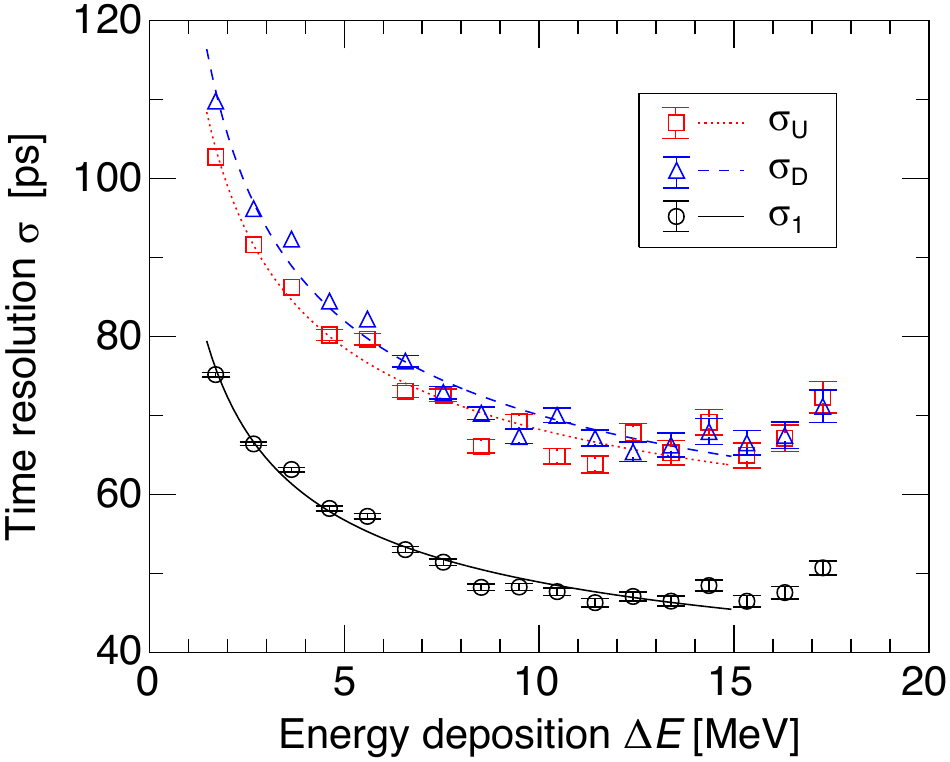}
\caption{ 
Dependence of time resolutions on the energy deposition $\Delta E$.
The dotted, dashed, and solid curves show fit results 
with $f(\Delta E) = p_0 + p_1/\sqrt{ \Delta E}$. Data sets corresponding to the 
first $0.1\times 10^{14}$ incident protons (the first data points in Figure~\ref{fig:time_resol_irr_time})
were used in this analysis.}
\label{fig:time_resol_dE_dep}
\end{figure}

The dependence of time resolution on energy deposition is presented 
in Figure~\ref{fig:time_resol_dE_dep}. 
The normalization of the energy deposition was performed by fitting the measured QDC–momentum correlation 
to a Monte Carlo simulation using the Geant4 framework \cite{Agostinelli_NIMA_2003_Geant4}.
The time resolution improves as the energy deposition increases, 
up to approximately 10~MeV, above which it remains nearly constant due to saturation.
Each time resolution plot can be fitted with an empirical function of the form 
$f(\Delta E) = p_0 + p_1/\sqrt{ \Delta E}$, as shown in the figure.

\begin{figure}[hbt]
\centering
\includegraphics[width=0.90 \linewidth]{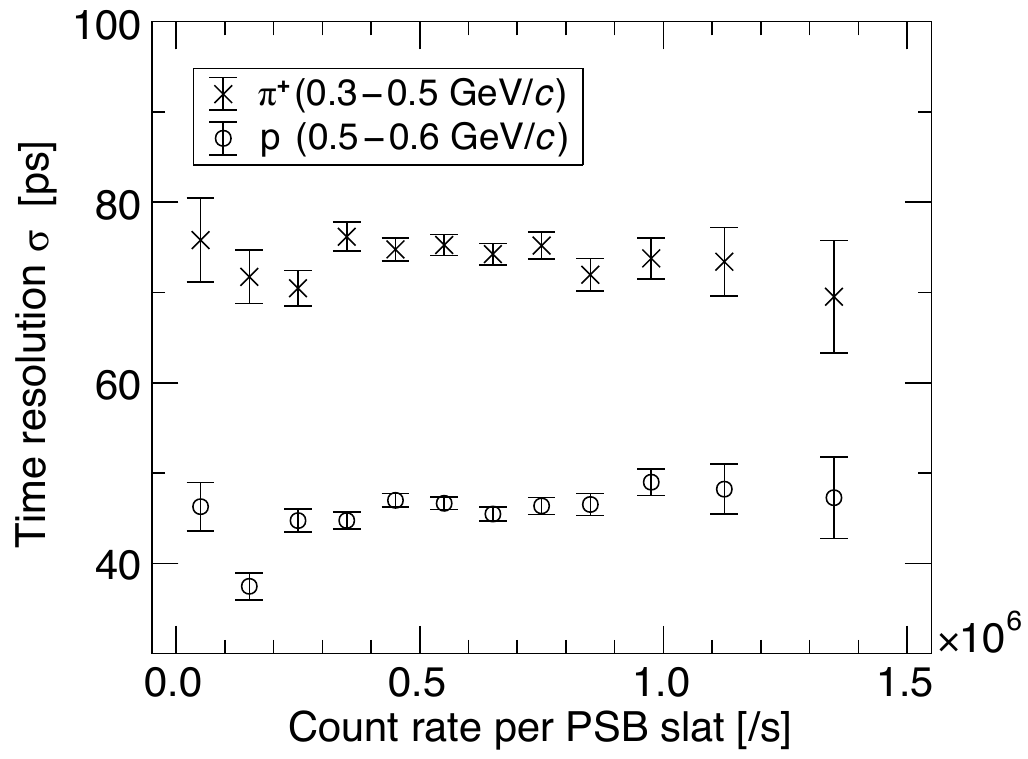}
\caption{ 
Dependence of time resolutions on the counting rate per individual PSB slat.}
\label{fig:time_resol_rate_dep}
\end{figure}

Figure~\ref{fig:time_resol_rate_dep} shows the analyzed time resolutions $\sigma_1$
as a function of the counting rate.
The time resolutions remain stable over the entire range of the count rate
for both the $\pi^{+}$ and proton data sets. 
No significant deterioration is observed up to a counting rate of 
$1.35 \times 10^6$~/s per single slat of the PSB. 
This demonstrates a substantial improvement in high-rate capability, 
also in terms of the time resolution, 
compared to the prototype previously reported in Ref.~\cite{Sekiya2022}.

\begin{figure}[hbt]
\centering
\includegraphics[width=0.92 \linewidth]{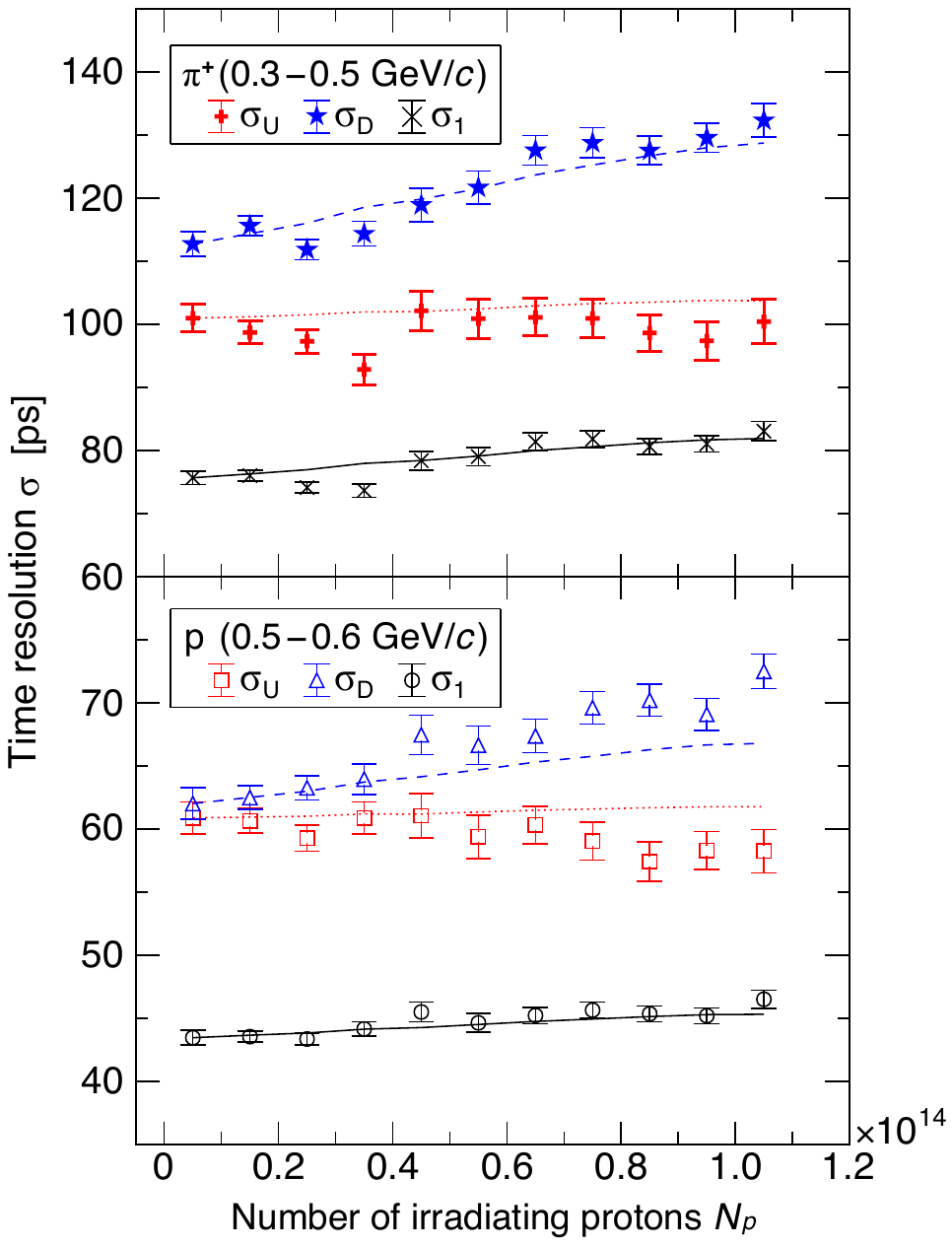}
\caption{ Evaluated time resolutions as functions of the total number of irradiating protons, $N_p$. The dotted, dashed, and solid curves present estimated trends, assuming that the time resolution was caused as a consequence of the QDC reduction.}
\label{fig:time_resol_irr_time}
\end{figure}

In Figure~\ref{fig:time_resol_irr_time}, the time resolutions are shown
as a function of the total number of protons impinging on the carbon target. 
The upper and lower panels correspond to the $\pi^{+}$ (0.3--0.5~GeV/$c$) 
and proton (0.5--0.6~GeV/$c$) data sets, respectively. 
The resolution per slat ($\sigma_1$) gradually deteriorates 
from 76~ps to 83~ps 
for the $\pi^{+}$ data, and from 43~ps to 46~ps for the proton data,
as the total number of the irradiating protons increases.
This deterioration is mainly attributed to the downstream MPPCs, 
as indicated by the decomposed resolution component ($\sigma_D$).

The observed increase of the time resolution can be interpreted
as a consequence of the reduced signal amplitude,
shown in Figure~\ref{fig:qdc_reducition_irr_time}, as follows. 
We assume and estimate the time resolution $\sigma_i (N_p) $ 
 after irradiation with $N_p$ protons  
 for each component $i = \mathrm{U}, \mathrm{D}, \mathrm{1}$
to be $\sigma_i(N_p) =  f_i( R_i(N_p) \cdot  f^{-1}_i(\sigma_{i}(n_0)))$.
Here, $n_0 = 0.05 \times 10^{14}$  is the averaged number 
of protons corresponding to the first data point 
in Figures~\ref{fig:qdc_reducition_irr_time} 
and \ref{fig:time_resol_irr_time}, 
$f_i$ is the fitted $\Delta E$ dependence of the time resolutions in Figure~\ref{fig:time_resol_dE_dep}, 
and $R_i(N_p)$ is the QDC reduction factor at $N_p$
relative to the first point at $n_0$ in Figure~\ref{fig:qdc_reducition_irr_time}.
The estimated trends for $\sigma_\mathrm{U}$, $\sigma_\mathrm{D}$, and $\sigma_1$ 
are presented in Figure~\ref{fig:time_resol_irr_time} by the dotted, dashed, and solid curves, respectively.
The observed increase in time resolution is well reproduced by this estimation, 
indicating that the degradation is primarily due to the signal amplitude reduction. 
This implies that no additional deterioration of the intrinsic time resolution of the MPPCs is observed 
as far as the present data are concerned. 

\section{Conclusion} \label{sec_conclusion}

In summary, the newly constructed plastic scintillator barrel with multiple-MPPC readout
was successfully operated in the first WASA-FRS experiment at GSI. 
Its performance, specifically the response to energy deposition and the time resolution,
was systematically investigated in terms of the dependence 
on the count rate and the total number of irradiating protons.
The time resolution was evaluated also as functions of the hit position and the energy deposition.

We observed a time resolution of approximately 75~ps
for minimum ionizing particles, which further improved to around 45~ps
with increasing the energy deposition. 
The detector maintained stable performance under high-rate conditions
up to 1.35~MHz per slat, with no significant deterioration in both the amplitude and timing response. 
Radiation-induced degradation of the signal amplitude was observed, particularly 
for MPPCs located near the reaction target,
with a reduction of approximately 35\% at an estimated 1~MeV neutron-equivalent fluence of $2.4 \times 10^{10}$~cm$^{-2}$. 
A slight deterioration in time resolution was also observed, which can be attributed to a consequence of the amplitude reduction.

These results demonstrate the overall performance of the MPPC-based plastic scintillator system 
under realistic experimental conditions with hadronic beams, 
including high counting rates and radiation exposure,
and would serve as a reference for the design and developments of future experiments 
with similar experimental conditions and requirements.

\vspace{5mm}
\noindent \textbf{Acknowledgements} \newline%

The authors would like to acknowledge the GSI staff for their support in the experiment. The WASA-FRS experiments were performed 
in the framework of the FAIR Phase-0 program at GSI.
This work is partly supported by JSPS Grants-in-Aid for Early-Career Scientists (Grant No.~JP20K14499) 
and for Scientific Research (B) (Grant No.~JP18H01242), 
JSPS Fostering Joint International Research (B) (Grant No.~JP20KK0070). 
The authors would like to acknowledge supports from the SciMat and qLife Priority Research Areas budget 
under the program Excellence Initiative-Research University at the Jagiellonian University, 
from Proyectos I+D+i 2020 (ref: PID2020-118009GA-I00), 
from the program ‘Atracci\'{o}n de Talento Investigador’ of the Community of Madrid (Grant 2019-T1/TIC-131), the Regional Government of Galicia under the Postdoctoral Fellowship Grant No.~ED481D-2021-018, the MCIN under Grant No.~RYC2021-031989-I, and from the European Union’s Horizon 2020 research and innovation programme (Grant No.~824093).

\end{document}